\documentclass[preprintnumbers,amsmath,amssymb,twocolumn,nopacs,prb]{revtex4}
\usepackage{graphicx}
\usepackage{pifont}
\usepackage{dcolumn}
\usepackage{bm}

\usepackage{amssymb}
\usepackage{amsmath}
\usepackage{array}

\newcommand{\carb}{$^{13}\textrm{C}$ }

\newcommand{\ket}[1]{\ensuremath{|#1\rangle}}
\newcommand{\bra}[1]{\ensuremath{\langle #1|}}

\begin{document}

\title{Coherent population trapping of a single nuclear spin \\
under ambient conditions}

\author{P. Jamonneau$^{1,\dag}$, G. H\'etet$^{1,2,\dag}$, A. Dr\'eau$^{1}$, J.-F. Roch$^{1}$, and V. Jacques$^{1,3}$}
\email{vjacques@ens-cachan.fr}
\affiliation{$^{1}$Laboratoire Aim\'{e} Cotton, CNRS, Universit\'{e} Paris-Sud and Ecole Normale Sup\'erieure de Cachan, 91405 Orsay, France}
\affiliation{$^{2}$Laboratoire Pierre Aigrain, CNRS, Universit\'e Pierre et Marie Curie, Universit\'e Paris Diderot and Ecole Normale Sup\'erieure, 75005 Paris, France}
\affiliation{$^{3}$ Laboratoire Charles Coulomb, Universit\'{e} de Montpellier and CNRS, 34095 Montpellier, France}
\altaffiliation{These authors contributed equally to this work.}

\begin{abstract}
Coherent control of quantum systems has far-reaching implications in quantum engineering. In this context, coherent population trapping (CPT) involving dark resonances~\cite{Gray,Arimondo_revue} has played a prominent role, leading to a wealth of major applications including laser cooling of atoms~\cite{Aspect} and molecules~\cite{Ni10102008}, optical magnetometry~\cite{Scully}, light storage~\cite{Lukin2000,Liu,Imamoglu_revue} and highly precise atomic clocks~\cite{Vanier_revue}. Extending CPT methods to individual solid-state quantum systems has been only achieved in cryogenic environments for electron spin impurities~\cite{Santori,Xu,Togan,Atature2014,Yale07052013,SiV,Pingault2014,Xia2015} and superconducting circuits~\cite{CPT_qubit1,CPT_qubit2,CPT_qubit,EIT_qubit}. Here, we demonstrate efficient CPT of a single nuclear spin in a room temperature solid. To this end, we make use of a three-level system with a $\Lambda$-configuration in the microwave domain, which consists of nuclear spin states addressed through their hyperfine coupling to the electron spin of a single nitrogen-vacancy defect in diamond. Dark state pumping requires a relaxation mechanism which, in atomic systems, is simply provided by spontaneous emission. In this work, the relaxation process is externally controlled through incoherent optical pumping and separated in time from consecutive coherent microwave excitations of the nuclear spin $\Lambda$-system. Such a pumping scheme with controlled relaxation allows us (i) to monitor the sequential accumulation of population into the dark state and (ii) to reach a new regime of CPT dynamics for which periodic arrays of dark resonances can be observed, owing to multiple constructive interferences. This work offers new prospects for quantum state preparation, information storage in hybrid quantum systems~\cite{Zhu,Kubo} and metrology.
\end{abstract}
\maketitle


\indent Nuclear spins in solids have attracted considerable interest over the last years owing to their high level of isolation from the environment, leading to very long coherence times~\cite{Morton2013,Maurer08062012,Sellars2015}. The detection and control of individual nuclear spins is usually achieved by exploiting their hyperfine coupling to an ancillary electronic spin. This can be performed by using a nitrogen-vacancy (NV) defect in diamond~\cite{Neumann06062008,Anais_PRL} whose spin triplet ground state ($S=1$) can be optically initialized, coherently manipulated with microwave fields and read-out by optical means through its spin-dependent photoluminescence (PL) intensity~\cite{Gruber1997}. Experiments making use of the hyperfine coupling between the NV defect and nearby long-lived nuclear spins are now reaching a level of control allowing to realize elaborate quantum information protocols~\cite{Waldherr,Pfaff01082014,Taminiau2014}. Here, we make use of electron spin transitions of a single NV defect to achieve coherent control of a single nuclear spin through CPT under ambient conditions. \\
\begin{figure}[h!]
\includegraphics[width = 8.3cm]{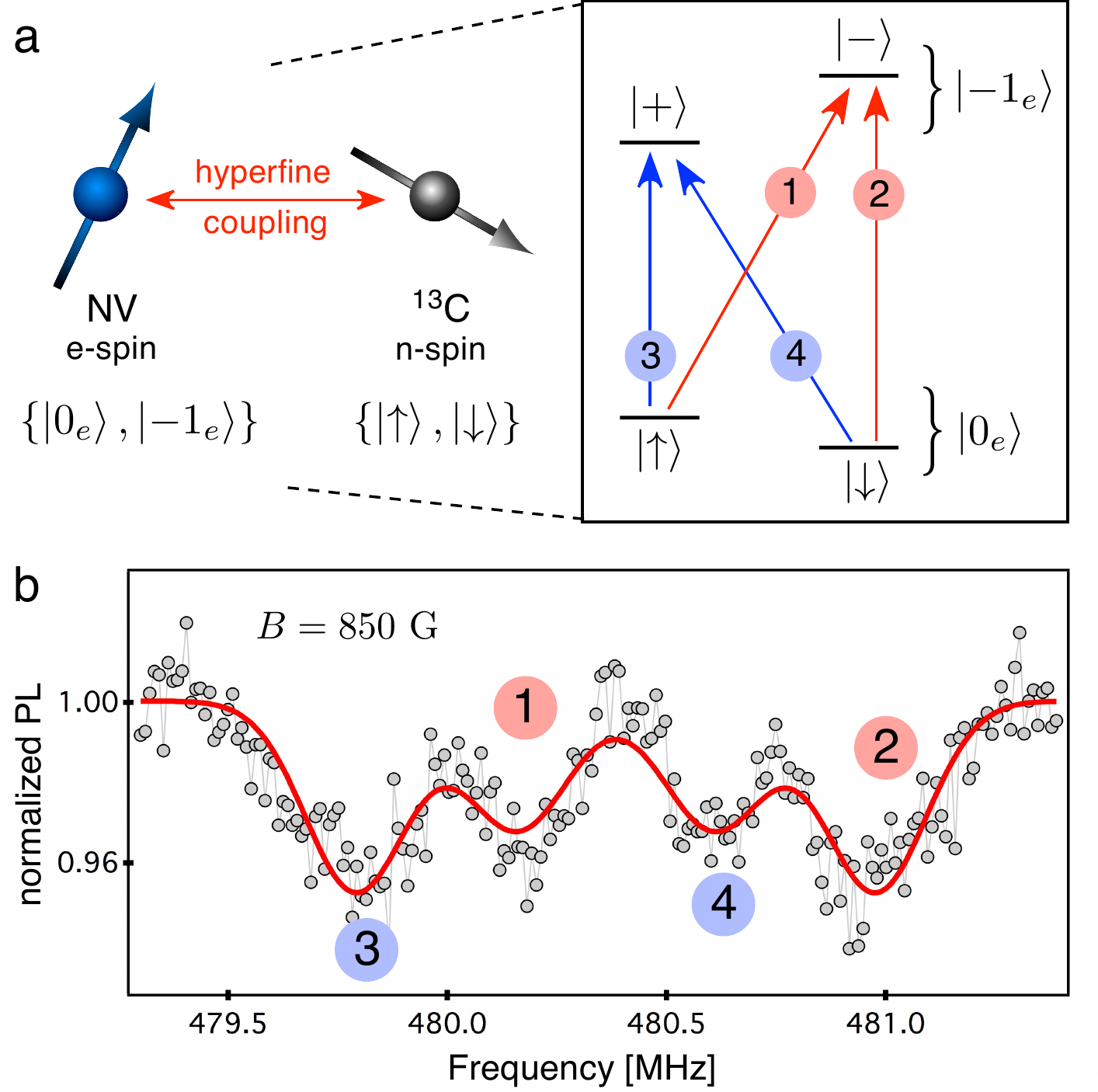}
\caption{{\bf Spin system}. {\bf a}, A single NV defect is coupled with a $^{13}$C nuclear spin via hyperfine interaction. The energy level structure shows the nuclear spin eigenstates in the $m_s=0$ ($\left|0_e \right.\rangle$) and $m_s=-1$ ($\left|-1_e \right.\rangle$) electron spin manifolds. The arrows indicate electron spin resonance (ESR) transitions. {\bf b}, ESR spectrum of the spin system recorded at a magnetic field $B=850$~G by sweeping the frequency of a microwave field while monitoring the spin-dependent PL intensity of the NV defect. Here the nuclear spin mixing parameter is $\theta\approx 1.14$~rad (see Supplementary Information).}
\label{Fig1}
\end{figure} 
\begin{figure*}[t]
\includegraphics[width = 16cm]{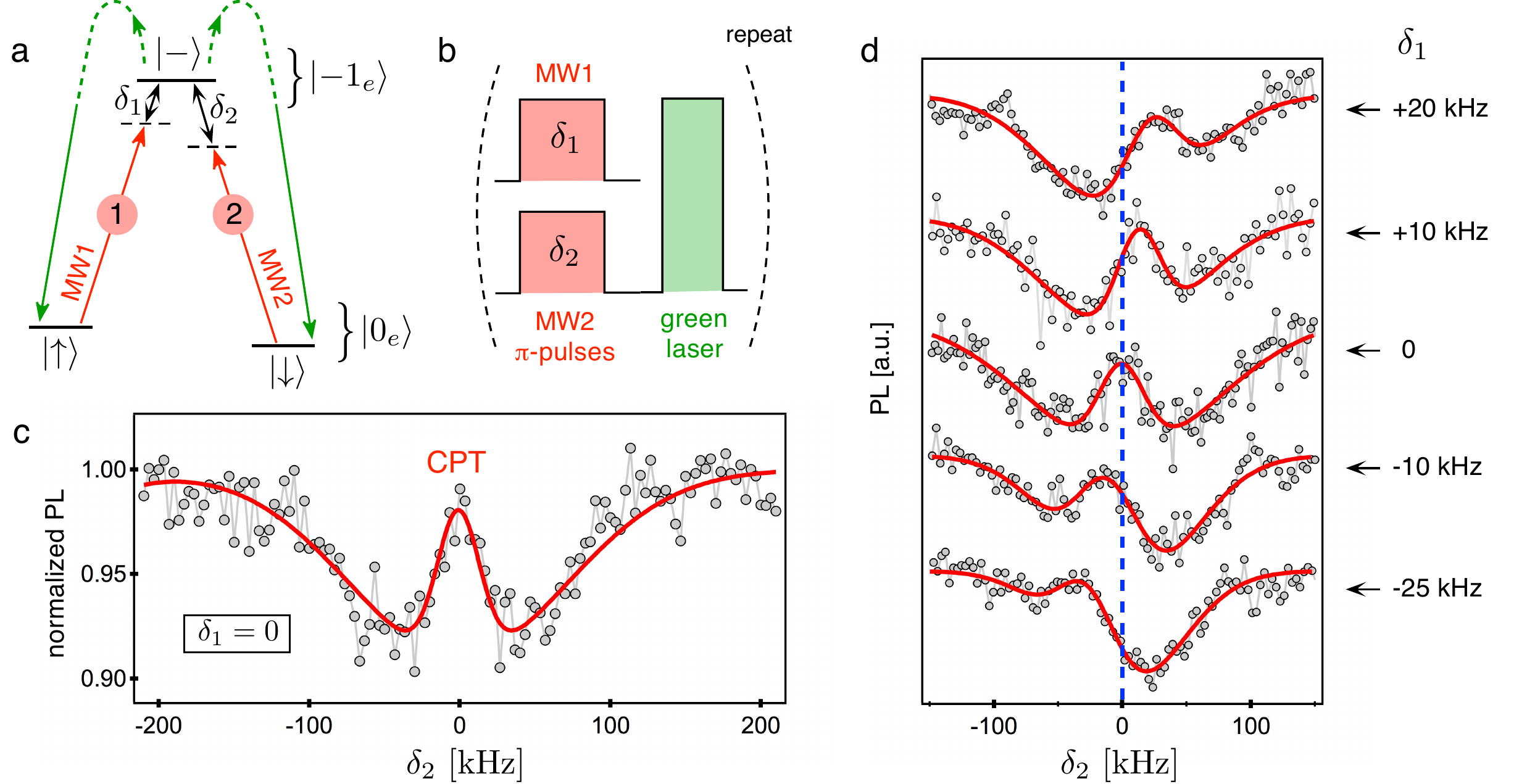}
\caption{{\bf Coherent population trapping of a single nuclear spin}. {\bf a,} $\Lambda$-scheme configuration. Relaxation from the excited state is triggered by optical pumping with a green laser. {\bf b,} CPT pulse sequence. Two selective MW $\pi$-pulses of duration $T_{\rm MW}=6 \ \mu$s are followed by a $300$-ns-long green laser pulse, with a power corresponding to the saturation of the NV defect optical transition. These parameters ensure spectral selectivity of an isolated $\Lambda$-scheme and enable optimizing the electronic spin readout contrast~\cite{Steiner2010}. {\bf c,} CPT dip observed in the ESR spectrum while setting $\delta_1=0$. For each sweep, the CPT sequence is repeated $N\approx 2000$ times before incrementing $\delta_2$. {\bf d,} ESR spectra recorded for various two-photon resonance conditions. The red solid lines are data fitting with two inverted Gaussian functions. All experiments are performed under ambient conditions with a magnetic field $B=850$~G applied along the NV axis. }
\label{coh}
\end{figure*}
\indent The spin system considered in this study is depicted in Fig.~\ref{Fig1}a. The electronic spin of a single NV defect is coupled by hyperfine interaction with a nearby $^{13}$C nuclear spin ($I=1/2$). In the $m_s=0$ ($\left|0_e \right.\rangle$) electron spin manifold, the nuclear spin eigenstates are denoted $\left|\uparrow \right.\rangle$ and $\left|\downarrow \right.\rangle$, corresponding to the nuclear spin projections along the NV defect quantization axis. In the $m_s=-1$ ($\left|-1_e \right.\rangle$) electron spin manifold, efficient nuclear spin mixing can be obtained by exploiting the anisotropic component of the hyperfine interaction in combination with a magnetic field applied along the NV defect axis~\cite{Anais_PRB}. The nuclear spin eigenstates are then given by
\begin{eqnarray*}
\left|+\right.\rangle&=&\cos\left(\frac{\theta}{2}\right)\left|\uparrow \right.\rangle+\sin\left(\frac{\theta}{2}\right)e^{i\phi}\left|\downarrow \right.\rangle\\
\left|-\right.\rangle&=&-\sin\left(\frac{\theta}{2}\right)e^{-i\phi}\left|\uparrow\right.\rangle+\cos\left(\frac{\theta}{2}\right)\left|\downarrow  \right.\rangle \ ,
\end{eqnarray*}
where the mixing parameter $\theta$ can be tuned with the magnetic field strength and the phase $\phi$ is fixed by the hyperfine tensor components [see Supplementary Information]. In this case, four transitions can be observed in the electron spin resonance (ESR) spectrum, with relative intensities fixed by the nuclear spin mixing parameter~\cite{Anais_PRB} [Fig.~\ref{Fig1}b]. The spin system therefore exhibits two $\Lambda$-level configurations in the microwave domain which can be used to perform CPT of the \carb nuclear spin. \\
\indent As illustrated in Fig.~\ref{coh}a, one of these $\Lambda$-schemes was isolated and its ESR transitions coherently excited with two microwave fields, MW1 and MW2, detuned from the resonances by $\delta_1$ and $\delta_2$. To briefly remind the general principle of CPT, we consider the two orthogonal linear superpositions of the lower levels  
\begin{eqnarray}
\left|B\right.\rangle&=&\frac{1}{\sqrt{\Omega_1^2+\Omega_2^2}}(\Omega_1\left|\uparrow \right.\rangle+\Omega_2\left|\downarrow \right.\rangle ) \\
\left|D\right.\rangle&=&\frac{1}{\sqrt{\Omega_1^2+\Omega_2^2}}\left(\Omega_2\left|\uparrow \right.\rangle-\Omega_1\left|\downarrow \right.\rangle \right)\ ,
\label{dark}
\end{eqnarray}
where $\Omega_1$ and $\Omega_2$ are the Rabi frequencies characterizing the strength of the MW coupling on each branch of the $\Lambda$-scheme. At the two-photon resonance condition $\delta_{\rm R}=\delta_1-\delta_2=0$, the population of the dark state $\left|D\right.\rangle$ is fully decoupled from the MW fields. On the other hand, the population of the bright state $\left|B\right.\rangle$ is efficiently coupled to the excited state, and then partially transferred to the dark state through spontaneous decay. After few excitation cycles, the system is thus trapped into the dark state. Such a CPT process requires an efficient relaxation from the excited state. For $\Lambda$-schemes in the optical domain, this is provided by the {\it irreversible} process of spontaneous emission~\cite{Arimondo_revue}. In this work, the intrinsic decay from the excited state $\left|-\right.\rangle$ is governed by the longitudinal electron spin relaxation of the NV defect. However, this process occurs within a typical timescale $T_1$ of few milliseconds at room temperature and is {\it bi-directional}, thus leading to a thermal state. To circumvent these detrimental effects, we use optical pumping to polarize the NV defect electronic spin in $\left|0_e \right.\rangle$. More precisely, the population transfer to the dark state is deterministically triggered within a sub-microsecond timescale by applying a green laser pulse. This is the key ingredient of our work. \\
\indent The experimental sequence used to observe CPT of a single nuclear spin is depicted in Fig.~\ref{coh}b. Two MW pulses driving selectively the two branches of the $\Lambda$-scheme are followed by a green laser pulse, which is used both for pumping the spin system into the dark state and for optical readout of the electronic spin state. ESR spectra were then recorded by continuously repeating this CPT sequence while monitoring the spin dependent PL intensity of the NV defect and sweeping the detuning $\delta_2$ across the resonance, with $\delta_1$ being fixed. A typical ESR spectrum recorded for $\delta_1=0$ is shown in Fig.~\ref{coh}c, revealing a pronounced CPT dip when the two-photon resonance condition is fulfilled. This is the signature of an efficient pumping of the spin system into the dark state. Remarkably, the contrast of the CPT dip is close to unity ($\approx 90 \%$), which implies that the laser pulse used to trigger the relaxation process is not altering significantly the nuclear spin coherence. Repeating the experiment for different values of $\delta_1$ shows that the position of the CPT dip shifts in accordance with the two-photon resonance condition $\delta_{\rm R}=0$ [Fig.~\ref{coh}d], confirming the dark state pumping interpretation.\\
\indent Such a pulsed CPT scheme gives a unique opportunity to analyze the sequential accumulation of population into the dark state~\cite{Soares} while increasing the number $N$ of repetitions of the CPT sequence. To this end, the two microwave fields were applied resonantly ($\delta_1=\delta_2=0$) and the probability $\mathcal{P}_{\left|D\right.\rangle}(N)$ to find the spin system in the dark state was inferred by measuring the population in state $\left|-\right.\rangle$ at each repetition of the CPT sequence through a calibrated spin-dependent PL intensity measurement (see Supplementary Informations). The experimental results are shown in Fig.~\ref{Fig3}a. Starting from a thermal state $\mathcal{P}_{\left|D\right.\rangle}(0)=1/2$, the dark state population increases exponentially with the number of CPT steps and saturates to $\mathcal{P}_{\left|D\right.\rangle}(\infty)\approx 0.88\pm0.03$, which corresponds to the dark state polarization efficiency. This dynamics is well reproduced by a simple rate equation model describing the competition between dark state pumping with a probability $\alpha_{p}$ per step and depolarization induced by nuclear spin decoherence with a probability $\alpha_{dp}$ per step. The probability $\alpha_{p}$ mainly depends on the MW pulse area~\cite{Soares} and the unbalanced relaxation between the two branches of the $\Lambda$-scheme, while the probability $\alpha_{dp}$ is essentially linked to nuclear spin decoherence induced by optical pumping (see Supplementary Information). Here, data fitting leads to $\alpha_{p}=0.43\pm0.07$, $\alpha_{dp}=0.12\pm0.03$ and the characteristic number of steps $N_s$ of the exponential growth is $N_s=1.4\pm0.3$. The dark state composition at the steady state was further analyzed by measuring $\left|\langle\downarrow|D\right.\rangle|^2$, {\it i.e.} the probability to find the spin system in state $\left|\downarrow\right.\rangle$, as a function of the Rabi frequencies $(\Omega_1,\Omega_2)$ of the microwave fields driving the $\Lambda$-system. As shown in Fig.~\ref{Fig3}b, this probability scales as $\Omega_1^2/(\Omega_1^2+\Omega_2^2)$, as expected from the dark state expression in the $\left\{\left|\uparrow \right.\rangle,\left|\downarrow \right.\rangle\right\}$ basis [see Eq.~(\ref{dark})]. This is a further evidence of CPT of the $^{13}$C nuclear spin. \\
\begin{figure}[t]
\includegraphics[width = 8.5cm]{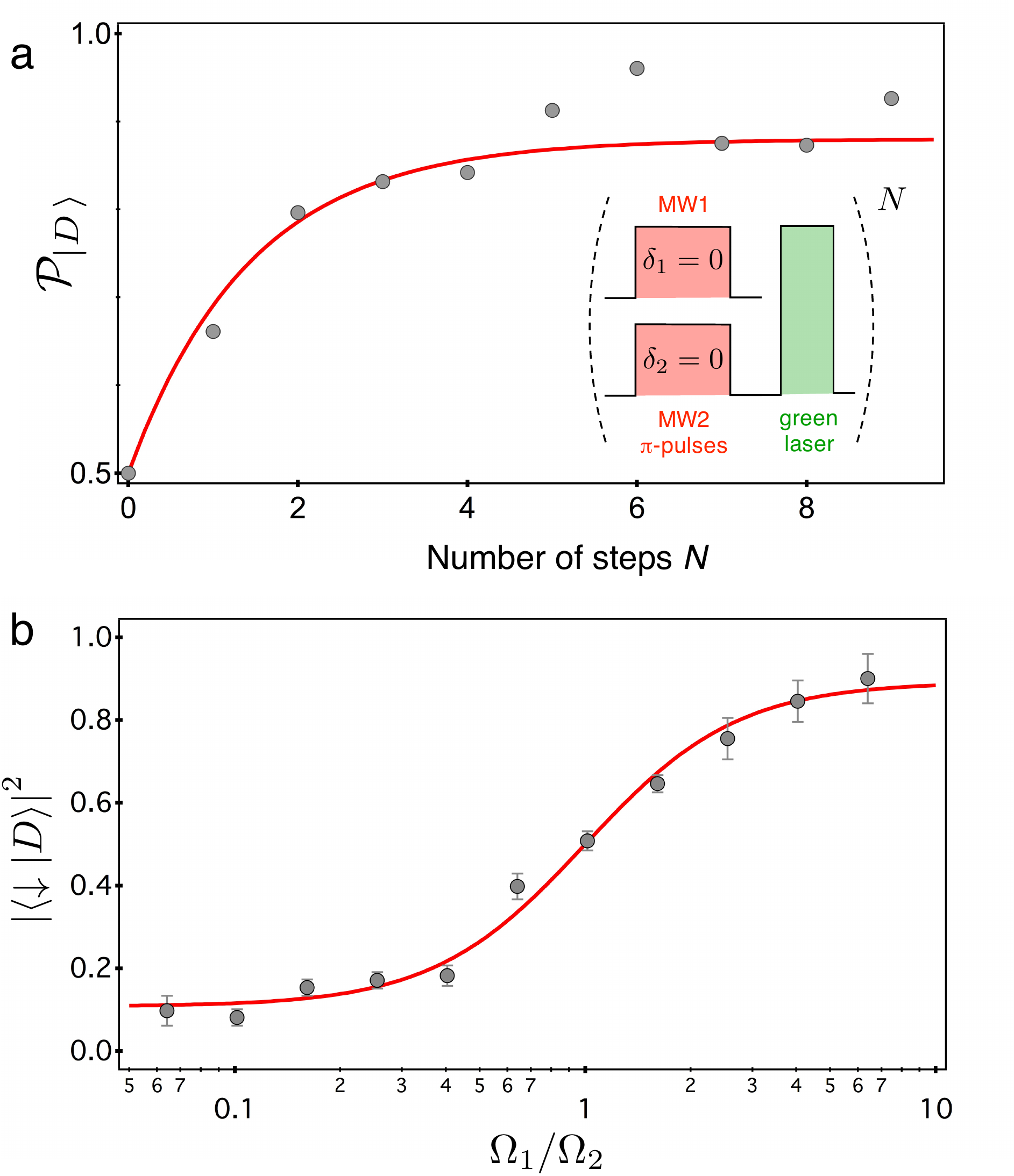}
\caption{{\bf Step-by-step dark state pumping}. {\bf a,} Probability $\mathcal{P}_{\left|D\right.\rangle}$ to find the nuclear spin in the dark state as a function of the number $N$ of CPT steps. The experiment is performed with $\Omega_1\approx \Omega_2$ using resonant microwave $\pi$-pulses. The solid line is data fitting with a rate equation model, leading to a characteristic number of steps $N_s$ of the exponential growth $N_s=1.4\pm0.3$. {\bf b,} Analysis of the dark state composition at the steady state ($N=20$) as the function of the Rabi frequencies ratio $r=\Omega_1/\Omega_2$. The solid line is data fitting with the function $r^2/(1+r^2)$ weighted by the imperfect CPT contrast (see Supplementary Information).}
\label{Fig3}
\end{figure}
\begin{figure*}[t]
\includegraphics[width = 18.3cm]{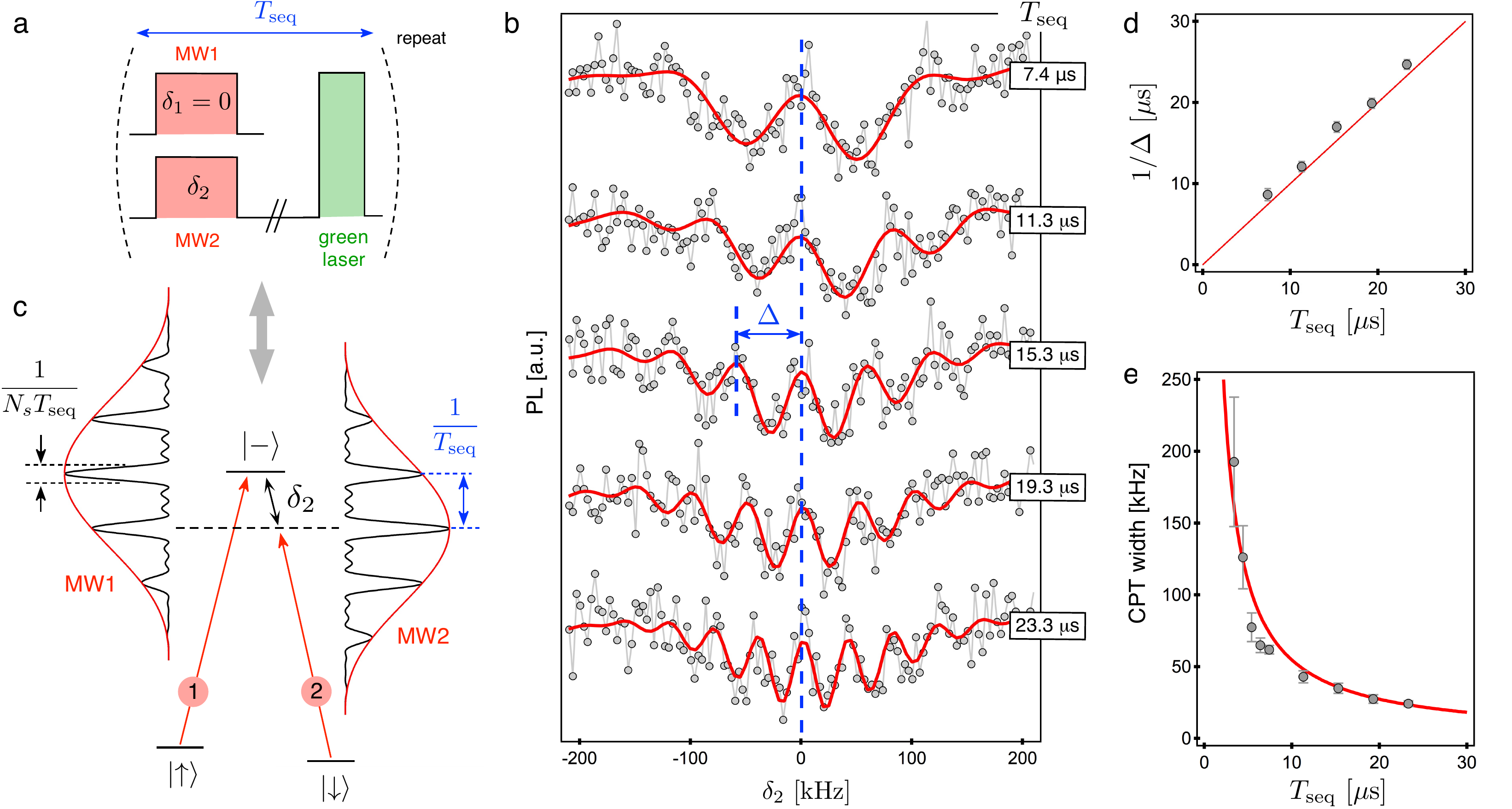}
\caption{{\bf Observation of multiple dark-resonances.} {\bf a,} The total duration of the CPT sequence $T_{\rm seq}$ is modified by varying the waiting time between coherent MW excitations and relaxation induced by optical pumping. {\bf b,} ESR spectra recorded in the same way as in Fig.~\ref{coh}c while increasing $T_{\rm seq}$ (from top to bottom). The red solid lines are Gaussian fits used to extract the periodicity and the width of the CPT dips. {\bf c,} Pulsed-MW excitations described as frequency combs driving the $\Lambda$-scheme. {\bf d,} Frequency of the dark resonance array $\Delta^{-1}$ as a function of $T_{\rm seq}$. The red solid line corresponds to $\Delta^{-1}=T_{\rm seq}$. {\bf e,} Full width at half maximum (FWHM) of the central CPT dip as a function of $T_{\rm seq}$. The red solid line is data fitting with the function $[N_sT_{\rm seq}]^{-1}$, leading to $N_s=1.8\pm 0.1$.}
\label{Fig4}
\end{figure*}
\indent As already pointed out, our pulsed-CPT experiment is performed in an unusual regime where the $\Lambda$-scheme relaxation is (i) deterministically triggered and (ii) well separated in time from the coherent excitation. These properties enable to precisely tune the time $T_{\rm seq}$ between subsequent coherent MW excitations [Fig.~\ref{Fig4}a] and analyze its effect on the properties of the dark resonance. For that purpose, ESR spectra similar to the one shown in Fig.~\ref{coh}c were recorded while increasing the duration of the CPT sequence. An array of CPT dips appears in the spectrum, whose period $\Delta$ evolves as $\Delta^{-1}=T_{\rm seq}$ [Fig.~\ref{Fig4}b,d]. Moreover, the width of the CPT resonances gets significantly narrower when $T_{\rm seq}$ increases [Fig.~\ref{Fig4}e]. To understand these observations, we first analyze the CPT pulse sequence in the spectral domain. As shown in Fig.~\ref{Fig4}c, the periodically repeated MW pulses of duration $T_{\rm MW}$ can be described by frequency combs, whose peaks are separated by $T_{\rm seq}^{-1}$ within a carrier-envelope of spectral width $T_{\rm MW}^{-1}$. In addition, the width of each peak of the comb is given by $[N_sT_{\rm seq}]^{-1}$, where $N_s$ is the characteristic number of CPT steps needed to reach the steady state of the spin system. Using this simple frequency comb picture, it is straightforward to show that dark state pumping is possible whenever the two-photon detuning is such that $\delta_{\rm R}=nT_{\rm seq}^{-1}$, where $n$ is an integer [Fig.~\ref{Fig4}c]. The spectral interval $\Delta$ between consecutive dark resonances then follows $\Delta^{-1}=T_{\rm seq}$, as experimentally observed [Fig.~\ref{Fig4}d]. Furthermore, the width of the CPT resonances closely follows the one of the comb peaks given by $[N_sT_{\rm seq}]^{-1}$ [Fig.~\ref{Fig4}e]. Data fitting leads to $N_s=1.8\pm 0.1$, in fair agreement with the step-by-step measurements reported in Fig.~\ref{Fig3}a.\\
\indent These multiple dark resonances can also be understood by considering the evolution of the dark state in the time-domain. For a two-photon detuning $\delta_{\rm R}$, it evolves as
\begin{equation}
\left|D(t)\right.\rangle=\frac{1}{\sqrt{\Omega_1^2+\Omega_2^2}}\left(\Omega_2\left|\uparrow \right.\rangle-\Omega_1 {\rm e}^{2i\pi\delta_{\rm R}t}\left|\downarrow \right.\rangle \right) \ , 
\end{equation}
corresponding to a precession at the frequency $\delta_{\rm R}$ between the states $\left|D\right.\rangle$ and $\left|B\right.\rangle$ defined by Eq.~(\ref{dark}). When the phase accumulated within $T_{\rm seq}$ is such that $\delta_{\rm R}=nT_{\rm seq}^{-1}$, the dark state is locked in phase with the periodic CPT pulse sequence, leading to efficient dark state pumping through multiple constructive interferences~\cite{Mlynek1981}. \\
\indent The dynamics of pulsed CPT with controlled relaxation markedly contrasts with continuous (CW) dark state pumping schemes. Indeed, CPT spectra recorded in CW regime usually result from a competition between the excited state relaxation rate $\gamma$ and the coherent coupling. This is manifested by the fact that the CPT width is given by $\Omega_2^2/\gamma$ in the limit where $\Omega_1\ll \Omega_2$ and neglecting decoherence of the ground state superposition~\cite{Bol91}. In this work, the relaxation is temporally decoupled from coherent coupling, and is not playing a role in the width of the resonance. In this novel regime, the width of the CPT dip is rather fixed by the sequence duration and $N_s$, which could be efficiently tuned by changing the area of the microwave pulses, as theoretically analyzed in Ref.~[\cite{Soares}].\\
\indent The fundamental limit to the CPT width is determined by two intrinsic parameters of the spin system. First, dark state resonances can be observed as long as $T_{\rm seq}$ is shorter than the longitudinal electron spin relaxation time $T_1$, which is on the order of few ms at room temperature, and can be extended to several tens of seconds at cryogenic temperature~\cite{Jarmola2012}. This limits the CPT width to $\Gamma_1=[N_sT_{1}]^{-1}$. Moreover, $T_{\rm seq}$ needs to be shorter than the dephasing time of the nuclear spin $T_{2n}^{*}=\Gamma_{2n}^{-1}$, which can be as long as few seconds under ambient conditions~\cite{Maurer08062012}. The fundamental limit to the CPT linewidth is therefore given by $\Gamma_0=\max\{\Gamma_1,\Gamma_{2n}^{*}\}$, which could potentially reach the sub-Hz regime at low temperature.

\indent In conclusion, we have demonstrated CPT of a single nuclear spin in a room temperature solid. The experiment is realized in a unique scenario for which dark state pumping is performed in a pulsed fashion while controlling the relaxation of a $\Lambda$-scheme in the microwave domain. This novel CPT regime may lead to new approaches for coherent control of individual nuclear spins, {\it e.g.} through stimulated Raman adiabiatic passage (STIRAP), information storage and processing in hybrid quantum systems~\cite{Zhu,Kubo} and precision atomic clocks~\cite{Vanier_revue}. 

We thank E. Arimondo, P. Pillet, D. Cl\'ement, P. Bertet, P. Maletinsky and J. R. Maze for fruitful discussions. This research has been partially funded by the French National Research Agency (ANR) through the projects A{\sc dvice}, Q{\sc invc} and S{\sc mequi}.

\begin{widetext}

\vspace{0.5cm}

\begin{center}
{\large SUPPLEMENTARY INFORMATION}
\end{center}
\section{Experimental setup}
\label{Exp}
We study native NV defects hosted in a commercial $[100]$-oriented high purity diamond crystal grown by chemical vapor deposition (Element6) with a natural abundance of $^{13}$C isotopes ($1.1\%$). Individual NV defects are optically isolated at room temperature using a home-built scanning confocal microscope under green laser excitation. Coherent manipulation of the NV defect electron spin is performed by applying a microwave field through a copper microwire directly spanned on the diamond surface. A permanent magnet mounted on a xyz-translation stage is used to apply a static magnetic field  along the NV defect axis. Details about the experimental setup can be found in Ref.~[\onlinecite{Anais_PRB2011}].

\section{Spin Hamiltonian}
\label{SpinSec}
The NV defect in diamond has a spin triplet ground state $S=1$, whose spin projection along the NV quantization axis ($z$) is denoted as $m_s=0,\pm1$. In this section, we do not introduce the hyperfine interaction between the NV defect electronic spin and its intrinsic $^{14}\rm{N}$ nuclear spin ($I_N=1$), since it is not playing any role in the reported experiments. We therefore consider a fixed $^{14}$N nuclear spin projection, {\it e.g.} $m_{I_{N}}=+1$. For a magnetic field amplitude $B$ applied along the NV defect axis, the ground-state spin Hamiltonian of the NV defect in frequency units reads 
\begin{eqnarray}\label{eq:h}
\hat{\mathcal{H}}_0=D\hat{S}_{z}^2+\gamma_eB\hat{S}_{z} \ , 
\end{eqnarray}
where $D \approx 2.87$ GHz is the zero-field splitting and $\gamma_e\approx2.8$~MHz/G is the electronic spin gyromagnetic ratio. When a neighbouring lattice site of the NV defect is occupied by a $^{13}$C isotope, corresponding to a nuclear spin $I=1/2$, the spin Hamiltonian becomes
\begin{eqnarray}\label{eq:h}
\hat{\mathcal{H}}=\hat{\mathcal{H}_{0}}+\gamma_{n}B\hat{I}_{z}+\mathbf{\hat{S}}\cdot{\mathcal A}\cdot\mathbf{\hat{I}} \ ,
 \end{eqnarray}
where $\gamma_n=1.07$~kHz/G is the gyromagnetic ratio of the $^{13}$C nuclear spin and $\mathcal{A}$ its hyperfine tensor. In the secular approximation, it simplifies as
\begin{eqnarray}\label{eq:hseq}
\hat{\mathcal{H}}=\hat{\mathcal{H}}_0+\gamma_nB\hat{I}_z+\hat{S}_z\mathcal{A}_{zz}\hat{I}_z+\frac{\mathcal{A}_{ani}}{2}\hat{S}_z(e^{-i\phi}\hat{I}_++e^{i\phi}\hat{I}_-) \ , 
\end{eqnarray}
where $\mathcal{A}_{ani}=(\mathcal{A}_{zx}^{2}+\mathcal{A}_{zy}^{2})^{1/2}$, $\tan\phi = \mathcal{A}_{zy}/\mathcal{A}_{zx}$ and $\hat{I}_{\pm}=\hat{I}_{x}\pm i\hat{I}_{y}$. 

\begin{figure}[t]
\begin{centering}
\includegraphics[width=10.5cm]{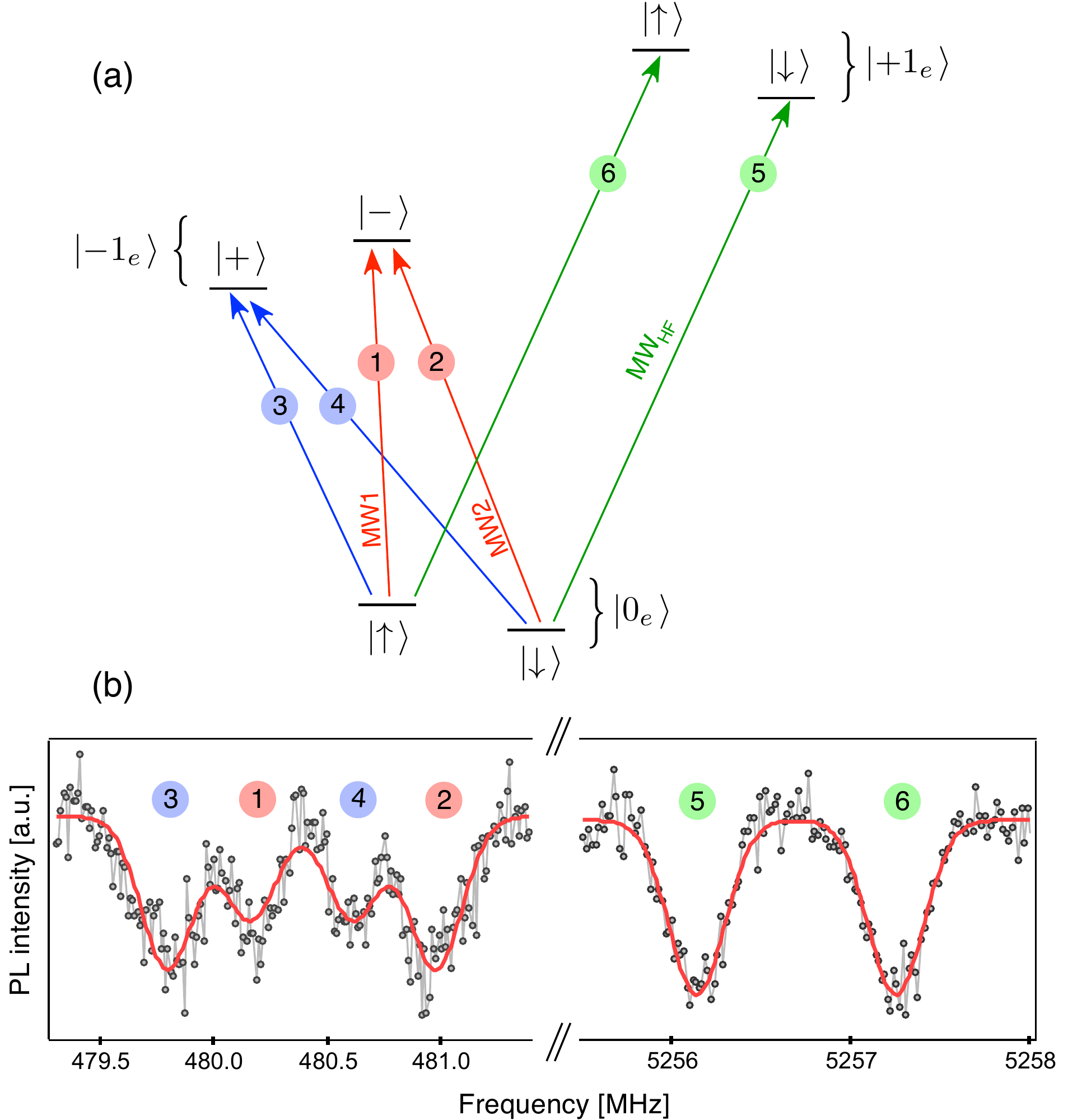}
\caption{(a)-Energy level structure of the spin system showing the nuclear spin eigenstates in the $m_s=0$ ($\ket{0_e}$) and $m_s=\pm1$ ($\ket{\pm1_e}$) electron spin manifolds. We consider $\mathcal{A}_{zz} \sim \gamma_nB$ and $\mathcal{A}_{zz}>0$. The colored arrows indicate electron spin transitions labelled from $\textcircled{1}$ to $\textcircled{6}$. (b)-ESR spectrum recorded for the coupled spin system with a magnetic field $B=850$~G applied along the NV defect axis. The low frequency part of the spectrum is reproduced from Fig.~1 of the main text.}
\label{Fig_ESR}
\end{centering}
\end{figure}

In the following we consider that $\mathcal{A}_{zz}>0$, and the eigenstates of the $\hat{I}_z$ operator are denoted as $\left|\uparrow \right.\rangle$ and $\left|\downarrow \right.\rangle$. As indicated in the main text, the eigenstates $\ket{\psi_i}$ of the spin system in the $m_s=0$ ($\left|0_e \right.\rangle$) and $m_s=-1$ ($\left|-1_e \right.\rangle$) electronic spin manifolds are given by~\cite{Anais_PRB}
\begin{eqnarray}\label{eq:eigenvec}
\ket{\psi_1}&=&\ket{0_e,\uparrow}\\
\ket{\psi_2}&=&\ket{0_e,\downarrow}\\
\ket{\psi_3}&=&\ket{-1_e,+}=\cos \left(\frac{\theta}{2}\right)\ket{-1_e,\uparrow}+\sin \left(\frac{\theta}{2}\right)e^{i\phi}\ket{-1_e,\downarrow}\\
\ket{\psi_4}&=&\ket{-1_e,-}=-\sin \left(\frac{\theta}{2}\right)e^{-i\phi}\ket{-1_e,\uparrow}+\cos \left(\frac{\theta}{2}\right)\ket{-1_e,\downarrow} \ ,
\end{eqnarray}
with
\begin{eqnarray}\label{eq:theta}
\tan \theta=\frac{\mathcal{A}_{ani}}{\mathcal{A}_{zz}-\gamma_nB} \ .
\end{eqnarray}

The corresponding eigenenergies read
\begin{eqnarray}\label{eq:eigenenergy}
\nu_1&=&\frac{\gamma_nB}{2}\\
\nu_2&=&-\frac{\gamma_nB}{2}\\
\nu_3&=&D-\gamma_eB-\frac{1}{2}\sqrt{\mathcal{A}_{ani}^2+(\mathcal{A}_{zz}-\gamma_nB)^2}\\
\nu_4&=&D-\gamma_eB+\frac{1}{2}\sqrt{\mathcal{A}_{ani}^2+(\mathcal{A}_{zz}-\gamma_nB)^2} \ .
\end{eqnarray}

When $\mathcal{A}_{zz} \sim \gamma_nB$, {\it i.e.} for $\theta \sim \pi /2$, four transitions from $m_s=0$ to $m_s=-1$ can be induced by a microwave field, owing to nuclear spin mixing in the $m_s=-1$ electronic spin manifold [see Fig.~\ref{Fig_ESR}]. Within this four level system, a $\Lambda$-scheme can be isolated, {\it e.g.} $\{\ket{0_e,\uparrow},\ket{0_e,\downarrow}, \ket{-1_e,-}\}$, in order to perform CPT of the $^{13}$C nuclear spin . 

For all the experiments reported in the main text, we use a nearby $^{13}$C nuclear spin whose hyperfine interaction is characterized by a longitudinal component $\mathcal{A}_{zz}\approx 1$~MHz and an anisotropic component $\mathcal{A}_{ani}=(\mathcal{A}_{zx}^{2}+\mathcal{A}_{zy}^{2})^{1/2}\approx0.3$~MHz. It corresponds to the lattice site denoted as I in Ref.~[\onlinecite{Anais_PRB}]. A magnetic field $B=850$~G is applied along the NV axis, which ensures efficient nuclear spin mixing in the $m_s=-1$ electronic spin manifold, while remaining far away from the ground-state level anti-crosssing ($B=1024$~G). At this field, the nuclear spin mixing is characterized by $\theta \approx1.14$~rad. The coherence time of the NV defect used in this work is  $T^*_2\approx 8$ $\mu \rm{s}$ at $B=850$~G, which is long enough to allow selective excitation of the $\Lambda$-system.

In the $m_s=+1$ ($\left|+1_e \right.\rangle$) electron spin manifold, which is not considered in the main text, the eigenstates of the spin system are given by 
\begin{eqnarray}\label{eq:eigenvec}
\ket{\psi_5}&=&\ket{+1_e,+^{\prime}}=\cos \left(\frac{\theta^{\prime}}{2}\right)\ket{+1_e,\downarrow}+\sin \left(\frac{\theta^{\prime}}{2}\right)e^{i\phi}\ket{+1_e,\uparrow}\\
\ket{\psi_6}&=&\ket{+1_e,-^{\prime}}=-\sin \left(\frac{\theta^{\prime}}{2}\right)e^{-i\phi}\ket{+1_e,\downarrow}+\cos \left(\frac{\theta^{\prime}}{2}\right)\ket{+1_e,\uparrow} \ ,
\end{eqnarray}
with
\begin{eqnarray}\label{eq:theta}
\tan \theta^{\prime}=\frac{\mathcal{A}_{ani}}{\mathcal{A}_{zz}+\gamma_nB} \ .
\end{eqnarray}

Here $(\mathcal{A}_{zz}+\gamma_nB) \gg \mathcal{A}_{ani}$, so that $\theta^{\prime}\approx 0$ and 
\begin{eqnarray}\label{eq:eigenvec}
\ket{\psi_5}&=&\ket{+1_e,\downarrow}\\
\ket{\psi_6}&=&\ket{+1_e,\uparrow} \ ,
\end{eqnarray}
with the corresponding eigenenergies 
\begin{eqnarray}\label{eq:eigenenergy}
\nu_5&=&D+\gamma_eB-\frac{1}{2}(\mathcal{A}_{zz}+\gamma_nB)\\
\nu_6&=&D+\gamma_eB+\frac{1}{2}(\mathcal{A}_{zz}+\gamma_nB) \ .
\end{eqnarray}

The $^{13}$C nuclear spin quantization axis is therefore parallel to the NV defect axis and only two nuclear-spin conserving transitions from $m_s=0$ to $m_s =+1$ can be observed in the ESR spectrum [Fig.~\ref{Fig_ESR}b].
These high-frequency ESR transitions were used to calibrate the experiments of sequential dark state pumping as explained in Section~\ref{SubSecExp}.

\section{Sequential dark state pumping}
\label{SecPump}

\subsection{Rate equation model}
\label{SubSecModel}

In this section, we analyze the experimental results reported in Figure 3 of the main text by introducing a simple rate equation model, which describes how the populations evolve within the $\Lambda$ system while repeating the CPT sequence. We consider $N$ successive dark state pumping steps with two microwave fields applied at resonance, $\delta_1=\delta_2=0$  [Fig. \ref{FigSI2}(a)]. The spin system is described in the basis $\{\ket{D},\ket{B}, \ket{-}\}$, where $\ket{D}$ and $\ket{B}$ are the dark and bright states of the $\Lambda$ system, which are defined by Eqs. (1) and (2) in the main text. Each pumping step can be decomposed into three different parts, as shown in Fig. \ref{FigSI2}(b).

\begin{enumerate}
\item{The transition $\ket{B}\rightarrow\ket{-}$ is first coherently driven though pulsed MW excitation with an effective Rabi frequency $\Omega = \sqrt{\Omega_1^2+\Omega_2^2}$ and a pulse area $A=\Omega\tau$, where $\tau$ is the MW pulse duration.}
 \item{The spin system is then kept in the dark during a waiting time shorter than the longitudinal relaxation time $T_1$ of the electronic spin. Here, the populations of the $\Lambda$ system do not evolve as long as the nuclear spin dephasing in the dark can be neglected.}
\item{A laser pulse is then applied at $t=T_{\rm laser}$, which triggers the relaxation from the excited state $\ket{-}$ to the ground states, $\ket{D}$ and $\ket{B}$, with rates $\Gamma_D$ and $\Gamma_B$ respectively. At the same time, the laser pulse induces nuclear spin dephasing with a rate $\Gamma_{dp}$. The laser pulse duration is denoted as $\Delta T= T_{\rm seq} - T_{\rm laser}$.}
\end{enumerate}

We first isolate the pumping step number $N$. The populations of the ground states at the beginning of step $N$ are equal to the ones obtained at the end of step $(N-1)$, which are labelled  $\mathcal{P}_{\ket{D}}(N-1)$ and $\mathcal{P}_{\ket{B}}(N-1)$ for state $\ket{D}$ and $\ket{B}$, respectively. Considering an ideal polarization of the NV defect electron spin under optical illumination, these populations are linked by the relation $\mathcal{P}_{\ket{D}}(N-1)+\mathcal{P}_{\ket{B}}(N-1) =1$. After coherent MW excitation, the population in state $\ket{-}$ is given by 
 \begin{equation}
 \mathcal{P}_{\ket{-}}(N)= \sin^2\left(\frac{A}{2}\right)\mathcal{P}_{\ket{B}}(N-1) = \sin^2\left(\frac{A}{2}\right)\left[1-\mathcal{P}_{\ket{D}}(N-1)\right] \ .
 \label{eq:initp-}
\end{equation}

\begin{figure}[t]
\begin{centering}
\includegraphics[width=11.5cm]{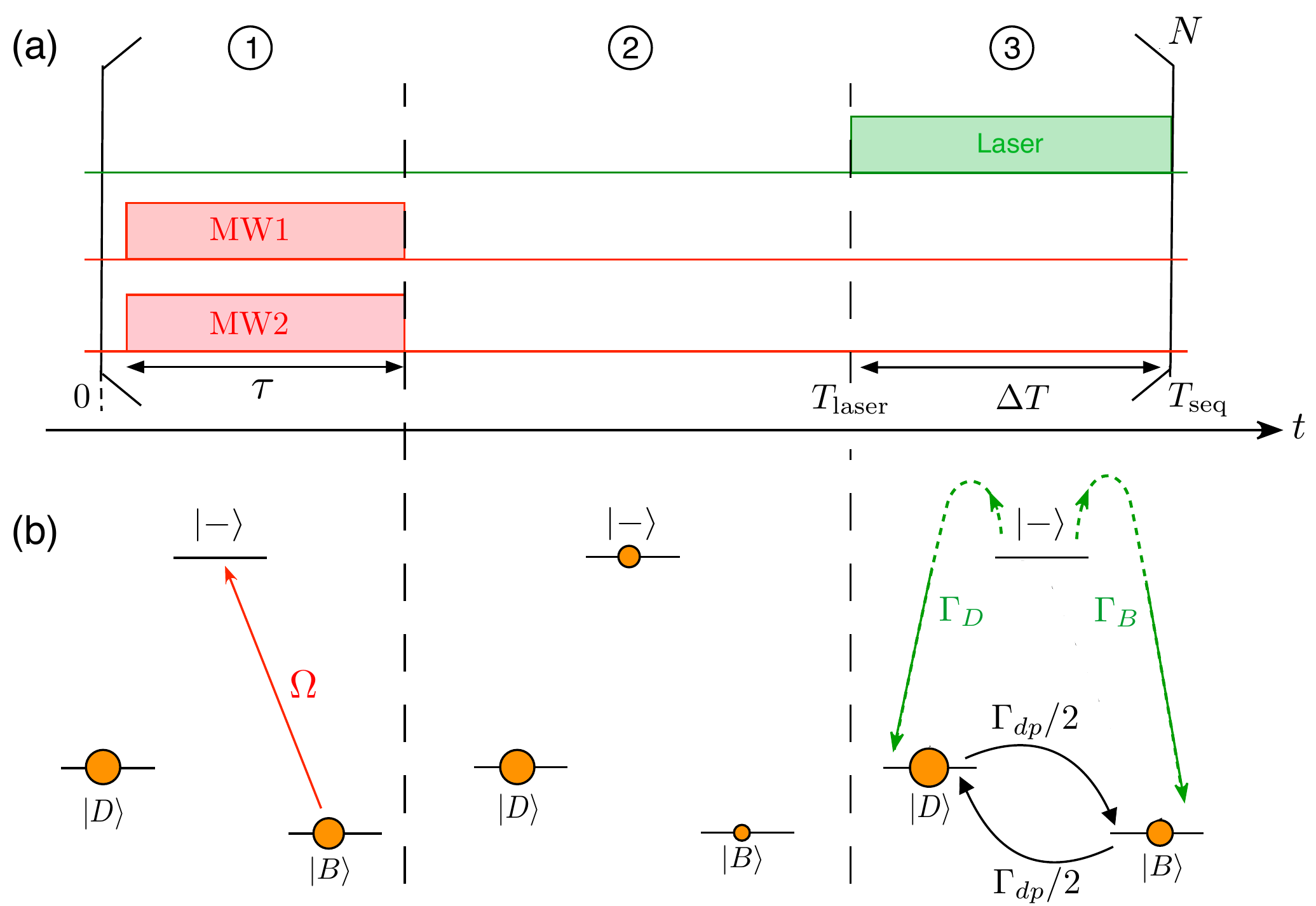}
\caption{(a)-Pulse sequence used for dark state pumping. (b)-Evolution of the populations in the $\Lambda$-system (represented as orange spheres) during the CPT sequence. $\Omega = \sqrt{\Omega_1^2+\Omega_2^2}$ is the effective Rabi frequency on the transition $\ket{B}\rightarrow\ket{-}$, $\Gamma_D$ and $\Gamma_B$ are the optically-induced decay rates and $\Gamma_{dp}$ is the dephasing rate of the nuclear spin under optical illumination.} 
\label{FigSI2}
\end{centering}
\end{figure}

During optical illumination, dark state pumping competes with laser-induced dephasing of the nuclear spin with a rate $\Gamma_{dp}$. We denote by $\alpha_p$ the probability to decay from $\ket{-}$ to the dark state at each CPT step. This probability is linked to the decay rates of the $\Lambda$ system through 
\begin{eqnarray}
\Gamma_D &=& \Gamma \alpha_p = \Gamma |\langle -|D \rangle |^2 \label{ap_def}\\
\Gamma_B &=& \Gamma (1-\alpha_p) = \Gamma |\langle -|B \rangle |^2 \ ,
\end{eqnarray}
where $\Gamma$ is the optically-induced polarization rate of the NV center electronic spin in the $m_s=0$ state~\cite{Anais_PRB2011}. 

The time evolution of the populations during the laser pulse, {\it i.e.} for $ \text{t} \in [T_{\text{laser}} , T_{\text{seq}}]$, is then described by the rate equations 
\begin{eqnarray}
\frac{d \mathcal{P}_{\ket{-}}}{dt} &=& \Gamma  \mathcal{P}_{\ket{-}} \label{eq:p-} \\
\frac{d \mathcal{P}_{\ket{D}}}{dt} &=& -\frac{\Gamma_{dp}}{2}( \mathcal{P}_{\ket{D}}- \mathcal{P}_{\ket{B}}) + \Gamma\alpha_p  \mathcal{P}_{\ket{-}}  \label{eq:pD}\\
\frac{d \mathcal{P}_{\ket{B}}}{dt} &=& \frac{\Gamma_{dp}}{2}( \mathcal{P}_{\ket{D}}- \mathcal{P}_{\ket{B}}) + \Gamma(1-\alpha_p)  \mathcal{P}_{\ket{-}} \ . \label{eq:pB}
\end{eqnarray}

Considering a closed $\Lambda$-system also implies that 
\begin{equation}
 P_{\ket{D}}(t) + P_{\ket{B}}(t) + P_{\ket{-}}(t) =1. \label{eq:pop}
\end{equation}

By solving equation (\ref{eq:p-}) for $ \text{t} \in [T_{\text{laser}} ,T_{\text{seq}}]$ with the initial condition at $\text{t} =T_{\text{laser}}$ given by Eq.(\ref{eq:initp-}), one can find 
\begin{equation}
\mathcal{P}_{\ket{-}}(t)  = \mathcal{P}_{\ket{-}}(T_{\text{laser}})e^{-\Gamma (t-T_{\text{laser}})} = \sin^2\left(\frac{A}{2}\right)\left[1-\mathcal{P}_{\ket{D}}(N-1)\right] e^{-\Gamma (t-T_{\text{laser}})}.
\label{eq:p-t}
\end{equation}

Using Eqs.~(\ref{eq:pop}) and (\ref{eq:p-t}), the evolution of the dark state population becomes
\begin{equation}
\frac{d\mathcal{P}_{\ket{D}}}{dt} + \Gamma_{dp}\mathcal{P}_{\ket{D}}(t) = \mathcal{P}_{\ket{-}}(T_{\text{laser}})e^{-\Gamma (t-T_{\text{laser}})}\left[\Gamma\alpha_p - \frac{\Gamma_{dp}}{2}\right] + \frac{\Gamma_{dp}}{2} \ ,
\end{equation}
leading to
\begin{equation}
\begin{split}
\mathcal{P}_{\ket{D}}(t) = \frac{\mathcal{P}_{\ket{-}}(T_{\text{laser}})}{\Gamma - \Gamma_{dp}}&\left[\Gamma\alpha_p - \frac{\Gamma_{dp}}{2}\right]\left[e^{-\Gamma_{dp} (t-T_{\text{laser}})} - e^{-\Gamma (t-T_{\text{laser}})}\right] \\
&+ \frac{1}{2}+\left[\mathcal{P}_{\ket{D}}(T_{\text{laser}})-\frac{1}{2}\right]e^{-\Gamma_{dp} (t-T_{\text{laser}})} \ ,
\end{split}
\end{equation}
where $\mathcal{P}_{\ket{D}}(T_{\text{laser}}) = \mathcal{P}_{\ket{D}}(N-1)$. 

At the end of the laser pulse of duration $\Delta T= T_{\rm seq} - T_{\rm laser}$, we consider that the electronic spin is fully polarized into $m_s=0$, so that $e^{-\Gamma \Delta T} \approx 0$. The link between the dark state population at the end of the CPT sequence $\mathcal{P}_{\ket{D}}(T_{\text{seq}}) = \mathcal{P}_{\ket{D}}(N)$ and the one at the beginning of the step  $\mathcal{P}_{\ket{D}}(N-1)$ then reads as 
\begin{equation}
\begin{split}
\mathcal{P}_{\ket{D}}(N) = \mathcal{P}_{\ket{D}}(N-1)e^{-\Gamma_{dp} \Delta T}\left[1-\frac{\sin^2\left(\frac{A}{2}\right)}{\Gamma - \Gamma_{dp}}[\Gamma\alpha_p - \frac{\Gamma_{dp}}{2}]\right] \\
+ \frac{1}{2}\left[1-e^{-\Gamma_{dp} \Delta T}\right] + \frac{\sin^2\left(\frac{A}{2}\right)}{\Gamma - \Gamma_{dp}}\left[\Gamma\alpha_p - \frac{\Gamma_{dp}}{2}\right] \ .
\end{split}
\end{equation}

By recurrence and with an initial thermal state of the nuclear spin [$\mathcal{P}_{\ket{D}}(\text{0})=0.5$], the dark state population after $N$ steps is finally expressed as 
\begin{equation}
\mathcal{P}_{\ket{D}}(N) = \frac{b}{1-a} + a^N\left[\mathcal{P}_{\ket{D}}(\text{0})-\frac{b}{1-a}\right] \ ,
\end{equation}
with 
\begin{eqnarray}
a &=& e^{-\Gamma_{dp} \Delta T}\left[1-\frac{\sin^2\left(\frac{A}{2}\right)}{\Gamma - \Gamma_{dp}}\left[\Gamma\alpha_p - \frac{\Gamma_{dp}}{2}\right]\right], \\
b &=& \frac{1}{2}\left[1-e^{-\Gamma_{dp} \Delta T}\right] + \frac{\sin^2\left(\frac{A}{2}\right)}{\Gamma - \Gamma_{dp}}\left[\Gamma\alpha_p - \frac{\Gamma_{dp}}{2}\right].
\end{eqnarray}

The characteristic number of steps needed to reach the stationary value of the dark state population is given by 
\begin{equation}
N_s=-1/\ln (a) = -1/\ln \left(e^{-\Gamma_{dp} \Delta T}\left[1-\frac{\sin^2\left(\frac{A}{2}\right)}{\Gamma - \Gamma_{dp}}\left[\Gamma\alpha_p - \frac{\Gamma_{dp}}{2}\right]\right]\right),
\end{equation}

and the dark state population at the steady state reads
\begin{equation}
\mathcal{P}_{\ket{D}}(\infty) = \frac{\frac{1}{2}\left[1-e^{-\Gamma_{dp} \Delta T}\right] + \frac{\sin^2\left(\frac{A}{2}\right)}{\Gamma - \Gamma_{dp}}\left[\Gamma\alpha_p - \frac{\Gamma_{dp}}{2}\right]}{1-e^{-\Gamma_{dp} \Delta T}\left[1-\frac{\sin^2\left(\frac{A}{2}\right)}{\Gamma - \Gamma_{dp}}\left[\Gamma\alpha_p - \frac{\Gamma_{dp}}{2}\right]\right]}.
\end{equation}

In order to simplify these equations, one can consider that dark state pumping is faster than laser-induced dephasing, {\it i.e.} $\alpha_p\Gamma \gg \Gamma_{dp}$. In our experiment, this assumption is supported by the observation of a large contrast of the CPT dip in the ESR spectrum, which is the signature of an efficient pumping of the spin system into the dark state [Fig. 2 of the main text]. Hence, the dark state population after $N$ pumping steps simplifies as 
\begin{equation}
\mathcal{P}_{\ket{D}}(N) =  \frac{\alpha_p'+\alpha_{dp}(0.5-\alpha_p')}{\alpha_p'+\alpha_{dp}(1-\alpha_p')}-0.5\frac{\alpha_p'(1-\alpha_{dp})}{\alpha_p'+\alpha_{dp}(1-\alpha_p')}\exp\left(- \frac{N}{ N_{s}}\right) \ ,
 \label{eq:pfit}
\end{equation}
where $\alpha_p' = \alpha_p \sin^2\left(\frac{A}{2}\right)$, $\alpha_{dp} = \left[1- \exp{(-\Gamma_{dp} \Delta T)}\right]$ is the depolarization probability per step induced by nuclear spin decoherence and $N_{s} = -1/\ln \left[1-\alpha_p'-\alpha_{dp}(1-\alpha_p')\right]$ is the characteristic number of pumping steps needed to reach the steady state of the $\Lambda$ system. At the steady state, the dark state population is then given by 
\begin{equation}
\mathcal{P}_{\ket{D}}(\infty) = \frac{\alpha_p'+\alpha_{dp}(0.5-\alpha_p')}{\alpha_p'+\alpha_{dp}(1-\alpha_p')} \ ,
\end{equation}
corresponding to a regime for which the gain by a pumping step into the dark state equals the losses induced by depolarization. If $\alpha_p' \gg \alpha_{dp}$, then $\mathcal{P}_{\ket{D}}(\infty) = 1$ and the system is fully polarized into the dark state. On the other hand, if $\alpha_{dp} \gg \alpha_{p}'$, then $\mathcal{P}_{\ket{D}}(\infty) = 0.5$ and the system remains in a thermal state. Hence, effective dark state pumping can be achieved as long as $\alpha_{p}' \ge \alpha_{dp}$. By fitting the step-by-step CPT experiments presented in Fig. 3 of the main text with equation (\ref{eq:pfit}), we obtain $\alpha_p' =0.43$ ($0.07$) and $\alpha_{dp} = 0.12$ ($0.03$), leading to $\mathcal{P}_{\ket{D}}(\infty) = 0.88$ ($0.03$).

\subsection{Polarization efficiency}
\label{SubSecPolar}

We now describe more precisely the polarization process. For clarity purpose, we assume that $\alpha_{dp} = 0$ and that the MW pulses bring the whole population of the bright state $\ket{B}$ to the excited state $\ket{-}$, {\it i.e.} $A=\pi$. We consider the asymmetric $\Lambda$ system represented in Figure \ref{FigSI2}(b). As a reminder, the spin states of the $\Lambda$-scheme are expressed by
\begin{eqnarray}
\ket{-} & = & -\sin \left(\frac{\theta}{2}\right)e^{-i\phi}\ket{\uparrow}+\cos \left(\frac{\theta}{2}\right)\ket{\downarrow} \label{state-} \\
\ket{D} & = &\frac{1}{\sqrt{\Omega_1^2 + \Omega_2^2}}\left(\Omega_2 e^{-i\psi}\ket{\uparrow} - \Omega_1 \ket{\downarrow}\right) \label{stateD}\\
\ket{B} & = &\frac{1}{\sqrt{\Omega_1^2 + \Omega_2^2}}\left(\Omega_1 \ket{\uparrow} + \Omega_2 e^{-i\psi} \ket{\downarrow}\right) \label{stateB}\ ,
\end{eqnarray}
where $\Omega_1$ and $\Omega_2$ are the Rabi frequencies of the two MWs driving the $\Lambda$-scheme. Here we have introduced the phase difference $\psi$ between the two MW fields, which was omitted in the main text. 

 The polarization efficiency is defined in Eq.(\ref{ap_def}) as $\alpha_p = |\langle -|D \rangle |^2$ corresponding to the probability to decay from $\ket{-}$ to the dark state during laser illumination. Using Eqs.~(\ref{state-}) and (\ref{stateD}), one can obtain 
\begin{equation}
\alpha_p = \frac{1}{1+(\Omega_1/\Omega_2)^2} \sin^2\left(\frac{\theta}{2}\right) + \frac{(\Omega_1/\Omega_2)^2}{1+(\Omega_1/\Omega_2)^2}\cos^2\left(\frac{\theta}{2}\right) + \frac{(\Omega_1/\Omega_2)}{1+(\Omega_1/\Omega_2)^2}\sin (\theta)\cos  (\phi - \psi) \ .
\label{ap}
\end{equation}

The parameters $\Omega_1$, $\Omega_2$ and $\psi$ allow for a full control over the dark state polarization efficiency and the $\Lambda$-scheme relaxation [see Fig.~\ref{Fig_theory_ap}]. This might enable to modify in a controlled fashion the dynamics of step-by-step dark state pumping.

\begin{figure}[h!]
\begin{centering}
\includegraphics[width=13cm]{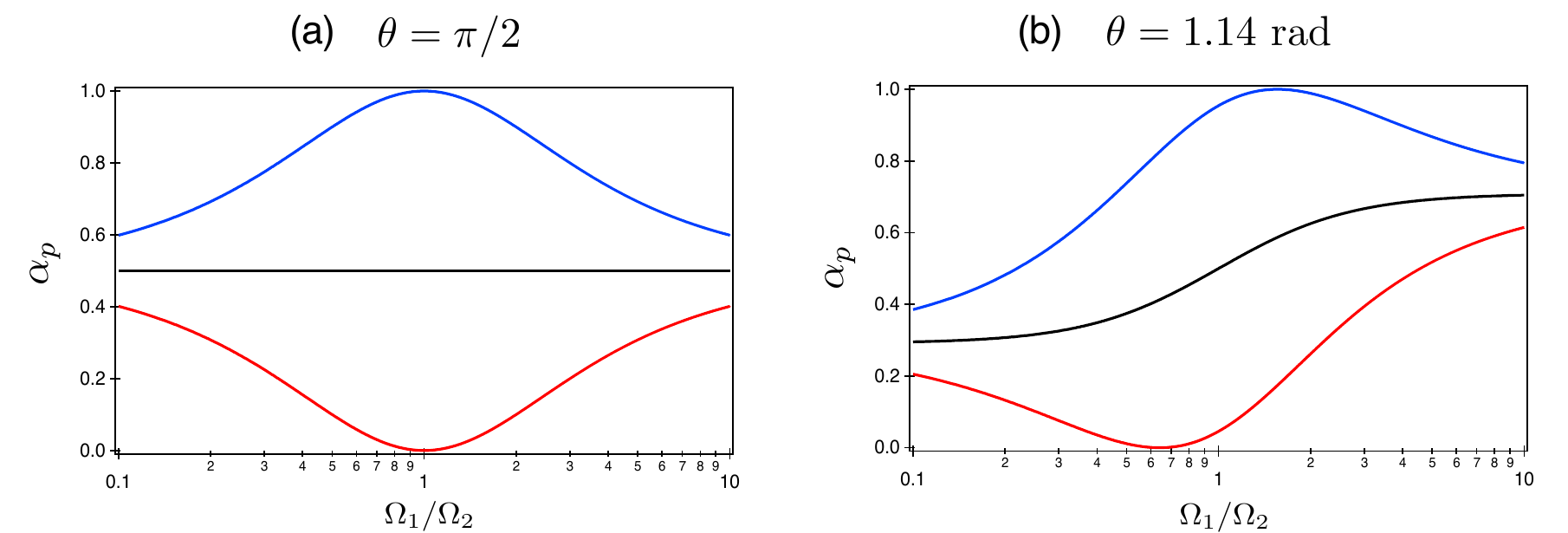}
\caption{Polarization coefficient calculated from Equation~(\ref{ap}) for (a) $\theta = \pi/2$ and (b) $\theta=1.14$~rad corresponding to the nuclear spin mixing parameter for a magnetic field $B=850$~G applied along the NV axis [see Section~\ref{SpinSec}]. In each case, the polarization coefficient is inferred for three different values of $\Delta \Phi =\phi - \psi $ ($\Delta \Phi = 0$ in blue, $\Delta \Phi = \pi/2$ in black and $\Delta \Phi = \pi$ in red).} 
\label{Fig_theory_ap}
\end{centering}
\end{figure}

\subsection{Experimental procedure}
\label{SubSecExp}

In this section, we give details of the experimental procedure which enables to extract the probability $\mathcal{P}_{\ket{D}}(N)$ through a calibrated spin dependent PL intensity measurement.\\

\noindent {\bf Reference sequence} - The high-frequency ESR transitions, presented at the end of Section~\ref{SpinSec},  were used to calibrate the experiments of sequential dark state pumping and to measure the nuclear spin's populations in $\left|\uparrow\right.\rangle$ and $\left|\downarrow\right.\rangle$. Assuming that the PL intensity is identical for $m_s=+1$ or $m_s=-1$, we use the sequence depicted on Fig.~\ref{Fig_ref}(a) to make a correspondence between PL intensity and population. The NV defect electronic spin state is first polarized in $m_s=0$ by laser illumination. After a $1$-$\mu \rm{s}$ long waiting time, a $300$-ns laser pulse is applied and the recorded PL signal is used as a reference signal for the $m_s=0$ spin sublevel. A selective MW $\pi$ pulse is then applied on the transition $\ket{0_e,\downarrow}\rightarrow\ket{+1_e,\downarrow}$ [MW$_{\rm HF}$ in Fig.~\ref{Fig_ESR}a], which brings half of the population into $m_s=+1$. A $300$-ns long laser pulse is then applied and the recorded PL signal is used as a reference signal for the $m_s=\pm 1$ spin sublevels. These measurements were used to calibrate the experiments of sequential dark state pumping.  \\

\noindent {\bf Step-by-step sequence} - As described in the previous paragraph, the experiment starts by applying a reference sequence which enables making the correspondence between PL intensity and populations of the spin system. Dark state pumping steps are then applied [Fig.~\ref{Fig_ref}(b)]. Each pumping step is composed by two $6$-$\mu \rm{s}$-long MW pulses driving the $\Lambda$-scheme at the two-photon resonance condition. Here the MW power is adjusted in order to obtain $\pi$-pulses, as verified by recording electron spin Rabi oscillations. After a $100$-ns-long waiting time, a $300$-ns laser pulse is used to pump the spin system into the dark state. The laser pulse is followed by a $1$-$\mu \rm{s}$ waiting time which ensures efficient relaxation of the population from the metastable state of the NV defect. The pumping steps are repeated $N$ times and the population in state $\ket{-}$ of the $\Lambda$-scheme, {\it i.e.} in $m_s=-1$,  is measured at each step by recording the spin-dependent PL during the laser pulses. After  $N$ repetitions, a $100$ $\mu \rm{s}$ laser pulse is used to reinitialize the system into a thermal state. The whole sequence is repeated continuously in order to improve the signal to noise ratio.\\

\begin{figure}[t]
\begin{centering}
\includegraphics[width=8cm]{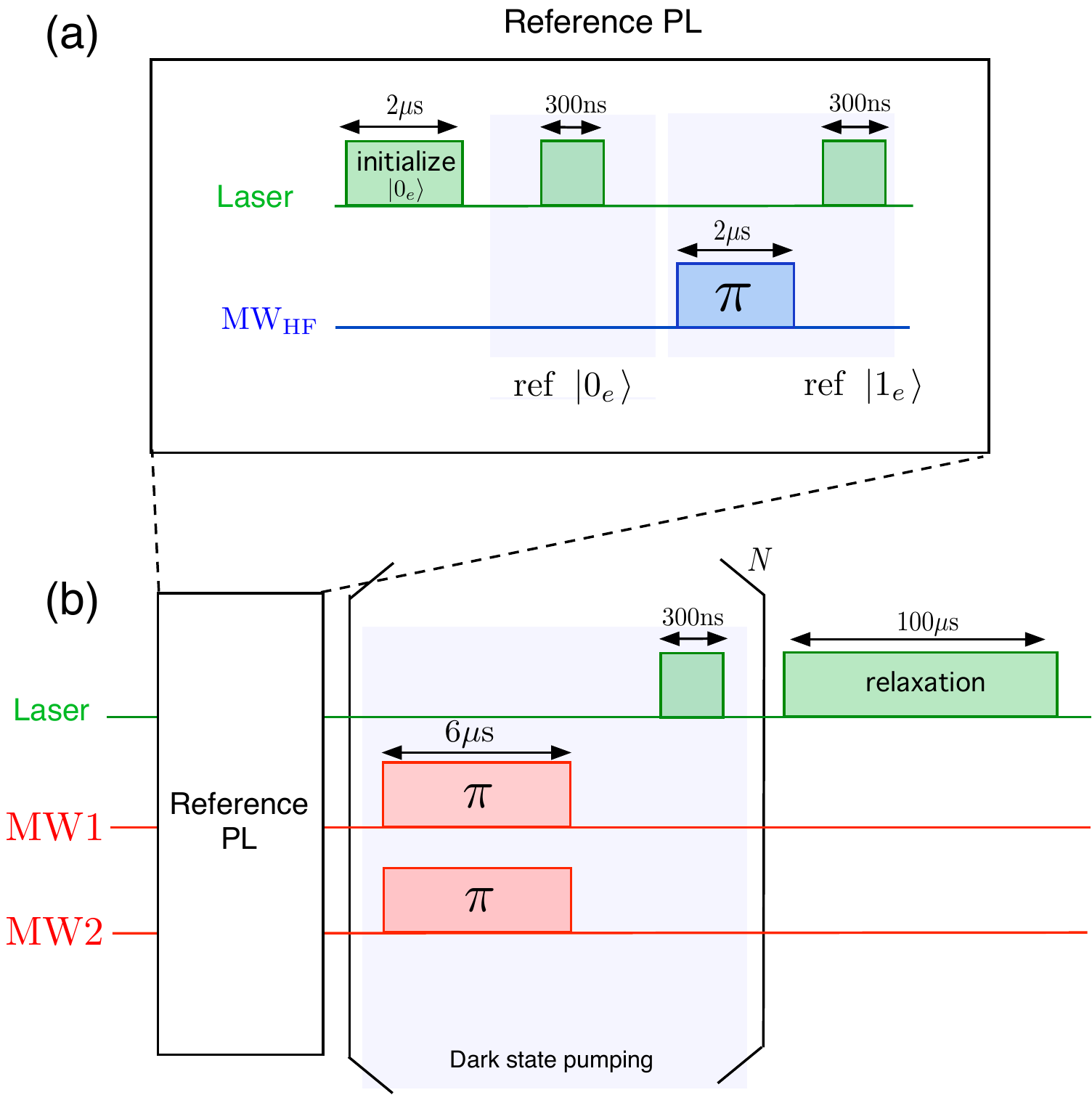}
\caption{(a)-Sequence used to make the correspondence between PL intensity and populations of the spin system. The microwave frequency $\text{MW}_{\text{HF}}$ is set on resonance with the transition $\ket{0_e,\downarrow}\rightarrow\ket{+1_e,\downarrow}$ [labeled $\textcircled{5}$ on figure \ref{Fig_ESR}(c)]. (b)-Experimental pulse sequence used for step-by-step dark state pumping.} 
\label{Fig_ref}
\end{centering}
\end{figure}

\noindent {\bf Extracting the dark state population} - The PL intensity measurement is done during each laser pulse of the CPT sequence. The recorded PL is compared to the reference signal to convert it in a measurement of the excited state $\ket{-}$ population, which follows Eq.~(\ref{eq:initp-}). Considering an initial thermal state of the nuclear spin [$\mathcal{P}_{\ket{D}}(0)=0.5$], we have
 \begin{equation}
\mathcal{P}_{\ket{-}}(1) = \frac{1}{2} \sin^2\left(\frac{A}{2}\right) \ .
\label{eq:FirstPop-}
\end{equation}

Using Eqs.~(\ref{eq:initp-}) and (\ref{eq:FirstPop-}), the probability to find the spin system in the dark state then follows 
\begin{equation}
\mathcal{P}_{\ket{D}}(N-1) = 1-\frac{\mathcal{P}_{\ket{-}}(N)}{2\mathcal{P}_{\ket{-}}(1)} .
\end{equation}

The sequential accumulation of population in the dark state can therefore be inferred by simply measuring the population in state $\ket{-}$, {\it i.e.} in $m_s=-1$, at each pumping step through a calibrated spin-dependent PL intensity measurement.

\subsection{Characterization of the dark state}
\label{SubSecCharac}

The experimental sequence used to characterize the dark state is sketched in Figure~\ref{FigSI5}. The dark state's composition is tuned by changing the power, and thus the Rabi frequencies ($\Omega_1$ and $\Omega_2$), of the MW fields driving the $\Lambda$ system. The pumping sequence is repeated $N = 20$ times in order to reach the steady state population of the dark state. The projection of the nuclear spin along the NV center axis is then measured by applying a selective $\pi$-pulse on the high-frequency ESR transition $\ket{0_e,\downarrow}\rightarrow\ket{+1_e,\downarrow}$ [MW$_{\rm HF}$ in Fig.~\ref{Fig_ESR}a]. The spin-dependent PL is measured by a $300$ ns laser pulse and gives the probability to find the nuclear spin in state $\left|\downarrow\right.\rangle$. A $100$ $\mu \rm{s}$-long laser pulse is finally applied in order to relax the system to a thermal state. The whole sequence is repeated continuously in order to improve the signal to noise ratio. \\

This measurement gives the probability to find the nuclear spin in state $\left|\downarrow\right.\rangle$. By representing the nuclear spin state with a density matrix $\rho_{nuc}$, the measured probability $\bra{\downarrow}\rho_{nuc}\ket{\downarrow}$ at the steady state is expressed as 
\begin{eqnarray}
\bra{\downarrow}\rho_{nuc}\ket{\downarrow} & = & \mathcal{P}_{\ket{D}}(\infty)|\bra{\downarrow}D\rangle |^2 + (1-\mathcal{P}_{\ket{D}}(\infty))|\bra{\downarrow}B\rangle |^2 \ .
\end{eqnarray}

Hence, one can find 
\begin{eqnarray}
|\bra{\downarrow}D\rangle |^2 & = & \frac{\bra{\downarrow}\rho_{nuc}\ket{\downarrow} - (1-\mathcal{P}_{\ket{D}}(\infty))}{2\mathcal{P}_{\ket{D}}(\infty)-1} \ .
\end{eqnarray}

\begin{figure}[t]
\begin{centering}
\includegraphics[width=13.5cm]{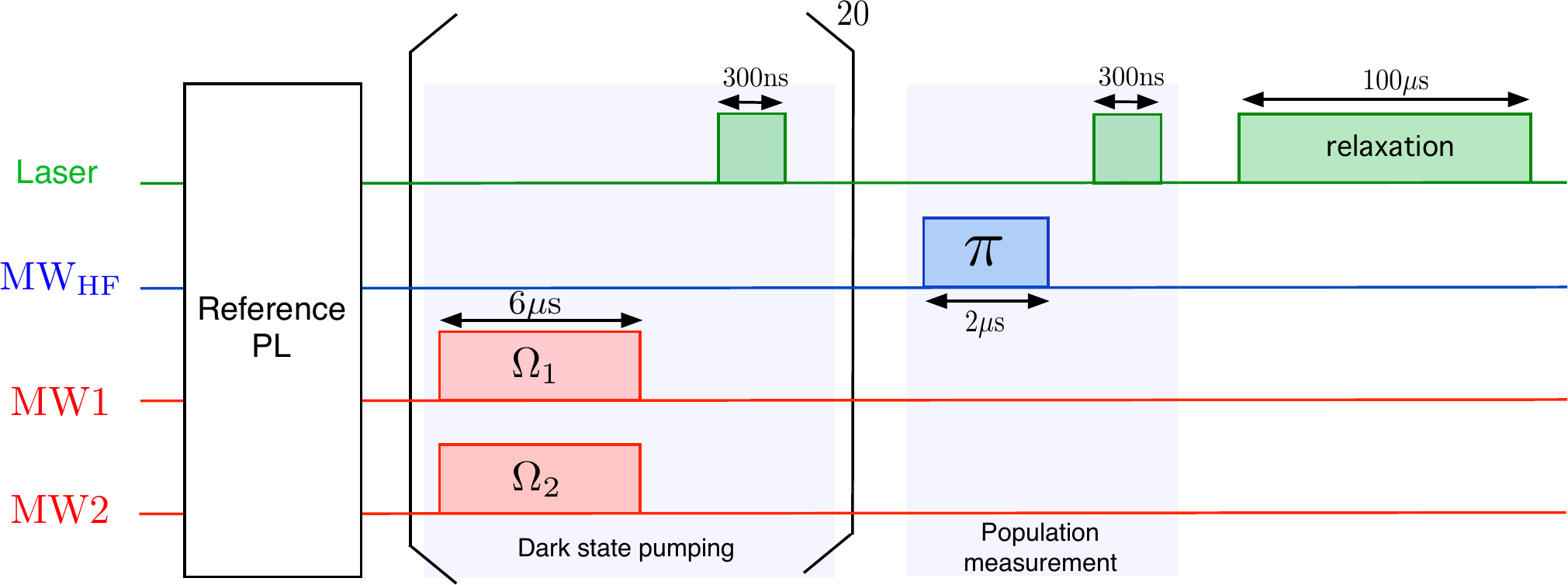}
\caption{Pulse sequence used for characterizing the dark state composition at the steady state.} 
\label{FigSI5}
\end{centering}
\end{figure}

In Figure 3(b) of the main paper, we plot $|\bra{\downarrow}D\rangle |^2 $ as a function of $\Omega_1/\Omega_2$. Ideally, when the system is in the dark state, the probability to find the nuclear spin in state $\left|\downarrow\right.\rangle$ is given by 
\begin{equation}
|\langle \downarrow | D\rangle |^2 = \frac{\Omega_1^2}{\Omega_1^2+\Omega_2^2} \ .
\label{dark_pop}
\end{equation}

To take into account the imperfections of the measurement, the experimental data were fitted with the function
\begin{equation}
f(\Omega_1/\Omega_2) = 0.5 + a\left(\frac{\Omega_1^2}{\Omega_1^2+\Omega_2^2}-0.5\right) \ ,
\end{equation}
where $a$ is an artificial contrast of the curve. Data fitting leads to $a=0.78\pm0.03$.

%
%
%
%


%
%
%
\bibliography{CPT}
\vspace{1 cm}
\end{widetext}

\end{document}